\documentclass[a4paper,11pt]{article}
\usepackage{pos}

\usepackage{booktabs}
\usepackage{cleveref}
\usepackage{dsfont}
\usepackage{siunitx}
\usepackage{physics}
\usepackage{subcaption}
\usepackage{todonotes}

\usepackage{soul}
\usepackage{xcolor}
\definecolor{darkteal}{RGB}{0,96,96}
\colorlet{darkteal}{darkteal!70!white}

\title{Real radiative decays of heavy pseudoscalar mesons}

\author*[1]{Teseo San Jose}
\author[2]{Yasumichi Aoki}
\author[3]{Matteo Di Carlo}
\author[3]{Felix Erben}
\author[1]{Vera G\"ulpers}
\author[1]{Maxwell T. Hansen}
\author[4]{Shoji Hashimoto}
\author[1]{Nils Hermansson-Truedsson}
\author[1]{Ryan Hill}
\author[4]{Takashi Kaneko}
\author[1]{Antonin Portelli}
\author[5]{Justus Tobias Tsang}

\affiliation[1]{School of Physics and Astronomy, University of Edinburgh, Edinburgh EH9 3FD, UK}
\affiliation[2]{RIKEN Center for Computational Science, Kobe 650-0047, Japan}
\affiliation[3]{Department of Theoretical Physics, CERN, 1211 Geneva 23, Switzerland}
\affiliation[4]{Institute of Particle and Nuclear Studiers, High Energy Accelerator Research Organization (KEK), Tsukuba 305-0801, JAPAN}
\affiliation[5]{Department of Mathematical Sciences, University of Liverpool, Liverpool L69 3BX, UK}

\emailAdd{teseo.sanjose@ed.ac.uk}

\abstract{We report our ongoing lattice QCD study of radiative leptonic decays of the charged pseudoscalar mesons $D$, $D_s$, $B$, and $B_c \to \ell \nu_\ell \gamma$. We carry out our analysis on a single JLQCD ensemble with lattice spacing $a=0.044~\text{fm}$. This work is a step towards a complete QCD+QED lattice calculation of these modes, aimed at reducing theoretical uncertainties in the extraction of $|V_{cd}|$ and $|V_{cs}|$ and providing first-principles estimates of the corresponding form factors in the $B$ sector.}

\FullConference{The 42nd International Symposium on Lattice Field Theory (LATTICE2025)\\
2-8 November 2025\\
Tata Institute of Fundamental Research, Mumbai, India\\}

\begin{document}
\maketitle

\section{Introduction}

\begin{figure}
    \centering
    \begin{subfigure}[t]{0.3\textwidth}
        \centering
        \includegraphics[scale=0.52]{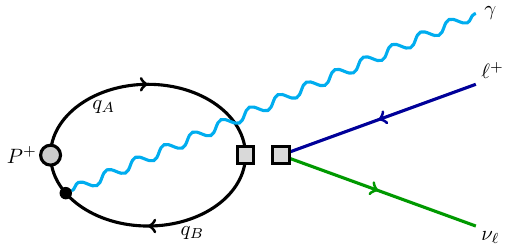}
        \caption{}
        \label{fig:quark-connected}
    \end{subfigure}
    \hfill
    \begin{subfigure}[t]{0.3\textwidth}
        \centering
        \includegraphics[scale=0.52]{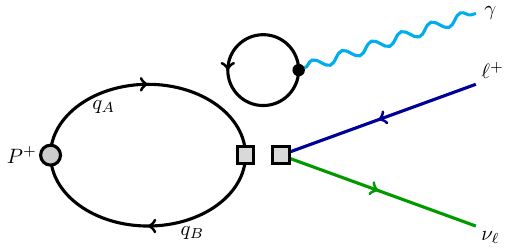}
        \caption{}
        \label{fig:quark-disconnected}
    \end{subfigure}
    \hfill
    \begin{subfigure}[t]{0.3\textwidth}
        \centering
        \includegraphics[scale=0.52]{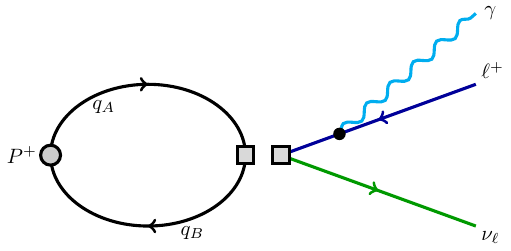}
        \caption{}
        \label{fig:lepton-emission}
    \end{subfigure}
    \caption{Diagrams contributing to the decay $P \to \ell \nu \gamma$ at leading order in $\alpha_{\text{em}}$. The double square indicates the four-fermion operator from Fermi theory. Here, we focus on the quark-connected contribution in \cref{fig:quark-connected}, where the photon can be emitted by either valence quark, and we plan to add the quark-disconnected contribution of \cref{fig:quark-disconnected} in the near future. The only non-perturbative contribution entering the final-state radiation in \cref{fig:lepton-emission} is the decay constant $f_P$.}
    \label{fig:diagrams}
\end{figure}

We study the radiative leptonic decays $P \to \ell \nu \gamma$ with an on-shell photon for the pseudoscalar mesons $P=D^{\pm}_{(s)},B^{\pm}_{(c)}$.
The contributions to leading order in $\alpha_\text{em}$ appear in \cref{fig:diagrams}; they are composed by two quark-connected pieces where the photon is emitted by either valence quark; a quark-disconnected contribution that we neglect at the moment but will include in the near future, and the final-state radiation.
These processes with very energetic photons give access to the meson's internal structure, which is encoded in the time-ordered hadronic tensor
\begin{equation}
    H_{V-A}^{\mu\nu}(k,p)=\int \text{d}^4 x~e^{ik \cdot x}  \mel{0}   {\text{T}\{J_{\text{em}}^\mu(x) J_{V-A}^\nu(0)\}} {P(p)},
\end{equation}
where $p$ and $k$ are the meson's and photon's 4-momenta, respectively, and we define the electromagnetic and weak currents as
\begin{equation}
    J_\text{em}^\mu(x) = \bar{q}(x) \gamma^\mu q(x), \qquad J_{V-A}^\nu(y)= \bar{q}_B(y)(\gamma^\nu-\gamma^\nu\gamma_5)q_A(y),
\end{equation}
with $q=q_A$ or $q=q_B$. By imposing charge conservation, $k_\mu H^{\mu\nu}=f_P p^\nu$, the hadronic tensor $H^{\mu\nu}$ can be decomposed into three terms, $H_{V-A}^{\mu\nu} = H_V^{\mu\nu} + H_{\text{pt}}^{\mu\nu} + H_A^{\mu\nu}$. Following the conventions in refs.~\cite{Bijnens:1992en,Carrasco:2015xwa,Beneke:2018wjp,PhysRevD.103.014502}, these can be Lorentz decomposed as
\begin{equation}
    \begin{split}
    H_V^{\mu\nu}(p,k) =& -i \epsilon^{\mu\nu\rho\sigma} k_\rho p_\sigma \frac{F_V(x_\gamma)}{m_P},
    \\
    H_{\text{pt}}^{\mu\nu}(p,k) =& f_P \left(g^{\mu\nu} + \frac{(2p-k)^\mu (p-k)^\nu}{2 p \cdot k - k^2}\right),
    \\
    H_A^{\mu\nu}(p,k) =& \frac{F_A(x_\gamma)}{m_P} \left[ g^{\mu\nu} (p \cdot k -k^2) - (p-k)^\mu k^\nu \right]
    \\
    &+\frac{H_1(x_\gamma)}{m_P} (g^{\mu\nu}k^2 -k^\mu k^\nu)
    \\
    &+\frac{H_2(x_\gamma)}{m_P} \frac{k^\mu(p\cdot k-k^2)-k^2(p-k)^\mu}{(p-k)^2-m_P^2}(p-k)^\nu.
    \end{split}
\end{equation}
The form factors $F_V$, $F_A$, $H_1$, and $H_2$ depend on the variable $x_\gamma = 2p \cdot k/m_P^2$, while $f_P$ is the decay constant and $m_P$ the meson mass. For a physical decay, $0 < x_\gamma < 1 - m_\ell^2/m_P^2$ with $m_\ell$ the charged lepton mass.
To compute the decay rate, we need the contraction $\epsilon^r_\mu H^{\mu\nu}$ of the hadronic tensor with a set of polarization vectors $\lbrace\epsilon^1_\mu,\epsilon^2_\mu\rbrace$; for an on-shell photon the relations
\begin{equation}
    k^2 = 0, \qquad \epsilon^r \cdot k = 0
\end{equation}
simplify the Lorentz decomposition substantially. In particular, the contributions from $H_1$ and $H_2$ vanish, leaving us with $F_V$ and $F_A$ for the real radiative decay,
\begin{multline}
    \epsilon_\mu^r H_{V-A}^{\mu\nu} = \epsilon_\mu^r \left( f_P \left[g^{\mu\nu} + \frac{p^\mu p^\nu - p^\mu k^\nu}{p \cdot k}\right]\right.
    \\
    \left.+ \left[ g^{\mu\nu} p\cdot k -p^\mu k^\nu \right] \frac{F_A(x_\gamma)}{m_P}
    -i \epsilon^{\mu\nu\rho\sigma} k_\rho p_\sigma \frac{F_V(x_\gamma)}{m_P}\right).
\end{multline}

In the literature, the current knowledge of these processes differs significantly between states. The RM123 collaboration were the first to provide $F_V$ and $F_A$ for $D$ and $D_s$ from the lattice \cite{PhysRevD.103.014502,PhysRevD.108.074505} in the electroquenched approximation, i.e., neglecting the diagram in \cref{fig:quark-disconnected}. Going to the heavier systems, HQET is the main theoretical framework used for $B$ decays, but even the most important coefficients parameterizing the hadron structure are poorly known \cite{aoife2025Slides,Beneke:2018wjp}.
Regarding experiment, there are only upper bounds for some radiative leptonic decays of $D_{(s)}$ and $B$, while $B_c$ remains unexplored \cite{ParticleDataGroup:2024cfk}, although the LHCb collaboration will study the latter for the first time to disentangle its signal from $B \to \mu\nu_\mu\gamma$ \cite{borsato2025Slides}.

Ultimately, real radiative decays can also be used to improve our understanding of other channels. For example, experimental cross sections of the purely leptonic decays $D_{(s)} \to \ell \nu_\ell$ are used, together with the meson decay constants $f_P$ obtained from the lattice, to extract the 2nd row of the CKM matrix at $\order{1\%}$ uncertainty \cite{ParticleDataGroup:2024cfk}. However, experiments can only provide the radiation inclusive process $P \to \ell\nu(\gamma)$, with both purely leptonic decays and processes with photons that are too soft to trigger the detector. Thus, one should consider the integration of all photon energies below that detector threshold. More generally, the most important isospin breaking effects need to be added to scrutinize just how reliable the current uncertainties really are.

In the remainder of these proceedings, we present the general method that is used to extract the form factors $F_V$ and $F_A$ from lattice simulations (in \cref{sec:methodology}), we detail our ongoing lattice simulation with domain-wall fermions (in \cref{sec:lattice_calculation}), and our initial dataset (in \cref{sec:dataset}). We finish outlining the next few steps in our project in \cref{sec:outlook}.

\section{Methodology}   \label{sec:methodology}

\begin{figure}[t]
    \centering
    \begin{subfigure}[t]{0.49\textwidth}
        \centering
        \includegraphics[scale=0.75]{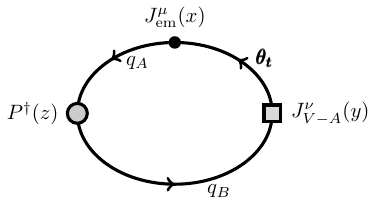}
    \end{subfigure}
    \hfill
    \begin{subfigure}[t]{0.49\textwidth}
        \centering
        \includegraphics[scale=0.75]{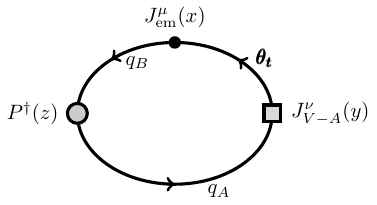}
    \end{subfigure}
    \caption{The two quark-connected diagrams that we compute, with the electromagnetic current inserted in either valence quark. To provide more photon momenta we introduce a twist angle $\theta_t$ in one quark line.}
    \label{fig:diagram-lattice}
\end{figure}

The form factors $F_V$ and $F_A$ for on-shell radiative decays can be extracted from the Euclidean correlation function \cite{PhysRevD.103.014502}
\begin{equation}
    \label{eq:3pt}
    C^{\mu\nu}_{3,V-A}(\pmb{p},\pmb{k},t_\gamma,t_W) = \int \text{d}^3x~e^{-i\pmb{k} \cdot \pmb{x}} \expval{J_\text{em}^\mu(\pmb{x},t_\gamma+t_W)J_{V-A}^\nu(\pmb{0},t_W) P^\dagger(\pmb{p},0)}
\end{equation}
where $t_W$ is the source-sink separation, the electromagnetic current is located at $t_\gamma+t_W$, and $P^\dagger(\pmb{p},0)$ creates a pseudoscalar meson with momentum $\pmb{p}$. To relate the correlator to the hadronic tensor $H^{\mu\nu}_{V-A}$, one needs to project to a particular photon energy $E_\gamma$, and form appropriate ratios comparing the spectral decomposition in Euclidean and Minkowski space \cite{Kane:2019jtj,Kane:2021zee,PhysRevD.103.014502,Giusti:2023pot,Giusti:2025ibe,DiPalma:2025iud},
\begin{equation}
    H^{\mu\nu}_{V-A}(\pmb{p},\pm{k}) = \lim_{t_W\to \infty} \lim_{T\to\infty} \frac{e^{m_P t_W} 2m_P}{\mel{P(\pmb{p})}{P^\dagger(0)}{0}} \int \text{d}t_\gamma e^{E_\gamma t_\gamma} C_{3,V-A}^{\mu\nu}(\pmb{p},\pmb{k},t_\gamma,t_W).
\end{equation}
In the integral, $t_\gamma \in [-t_W, t_\text{max}]$ with $t_\text{max} \ll T-t_W$ to avoid unphysical time orderings \cite{DiPalma:2025iud}.
Note that the hadronic tensor as extracted from the lattice depends on the source-sink separation $t_W$ and the time extent $T$, which need to be taken to infinity to extract the physical result. Also, different $(\mu,\nu)$ components and momentum configurations may exhibit different lattice artifacts. In order to study these effects systematically, we compute all non-zero components of the $C_3^{\mu\nu}$ correlator for every value of $t_W$ as explained in \cref{sec:lattice_calculation}.

\section{Lattice calculation} \label{sec:lattice_calculation}

\begin{figure}[t]
    \centering
    \begin{subfigure}[t]{0.45\textwidth}
        \centering
        \includegraphics[scale=1.3]{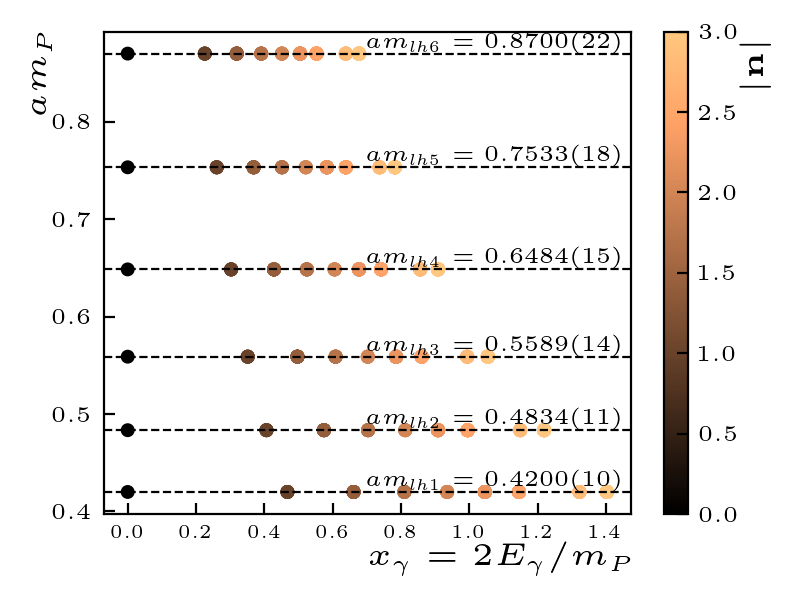}
        \caption{}
        \label{fig:fourier-modes}
    \end{subfigure}
    \begin{subfigure}[t]{0.45\textwidth}
        \centering
        \includegraphics[scale=1.3]{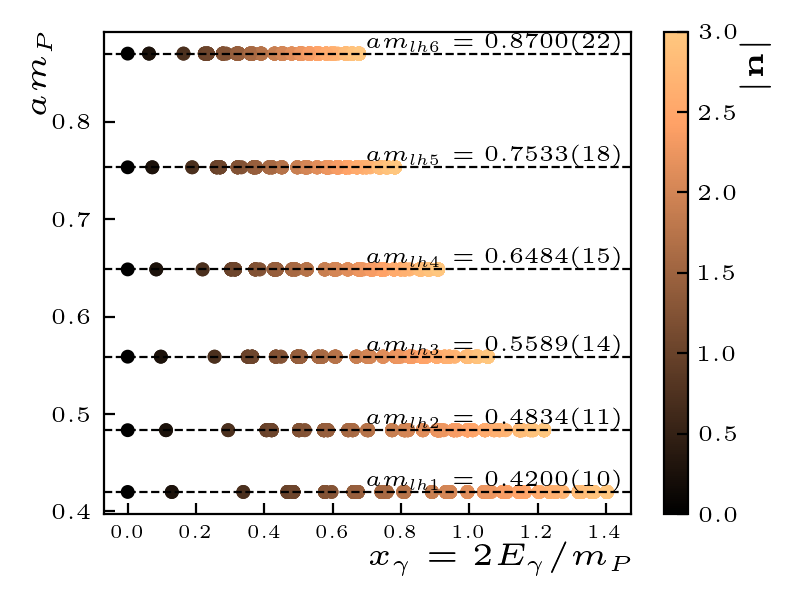}
        \caption{}
        \label{fig:twist-modes}
    \end{subfigure}
    \caption{The six heavy-light pseudoscalar meson masses vs the value of the Lorentz invariant $x_\gamma$ in the center-of-mass frame, with the color indicating the photon Fourier mode. \Cref{fig:fourier-modes} shows the values of $x_\gamma$ that we can access with our methodology relying only on Fourier modes. As shown in \cref{fig:twist-modes}, we use twisted boundary conditions to cover additional values of $x_\gamma$. The physical region corresponds to $0 < x_\gamma < 1 - m_\ell^2/m_P^2$, where $m_\ell$ and $m_P$ are the lepton and pseudoscalar meson masses.}
    \label{fig:modes}
\end{figure}

\begin{table}[t]
    \centering
    \begin{tabular}{cccc cccc}
        \toprule
        $\beta$ &
        $a^{-1}~[\unit{\giga\eV}]$ &
        $L^3 \times T \times L_s$ &
        $am_{ud}$ &
        $am_s$ &
        $m_{\pi}[\unit{\mega\eV}]$ &
        $m_{\eta_s}[\unit{\mega\eV}]$ &
        $m_{\pi}L$ \\
        \midrule
        4.47 & \num{4.496(80)} & $64^3 \times 128 \times 8$ & 0.003 & 0.015 & 284 & 627 & 4.0 \\
        \bottomrule
    \end{tabular}
    \caption{JLQCD ensemble used in this work \cite{PhysRevD.109.074503}. From left to right, bare coupling, lattice spacing, geometry of the 5D fields, light and strange bare-quark masses, and light and strange pseudoscalar meson masses.}
    \label{tab:ensembles}
\end{table}

We carry out our lattice simulations on the JLQCD ensemble shown in \cref{tab:ensembles}.
The gauge configurations are distributed according to a tree-level improved Symanzik action and saved to disk every 100 MDU.
The links are stout-smeared \cite{PhysRevD.69.054501,Cho:2015ffa} three times with weight $\rho=0.1$.
For the fermion sector, we employ a M\"obius domain wall action \cite{Brower:2012vk} with $b+c=2$, $b-c=1$, and a Wilson-Dirac operator as kernel with $M=1$. For more details on the ensemble, see Ref.~\cite{PhysRevD.109.074503}.
The $ud$ and $s$ bare quark masses appear in \cref{tab:ensembles}, and are the same for the valence and sea sectors.
To study the $D$ and $D_s$ mesons, we tune the valence charm-quark mass to reproduce the PDG value of the charmonium spin-averaged mass,
$(m_{P_{cc}} + 3m_{V_{cc}})/4$ where $P_{hh}$ and $V_{hh}$ are the pseudoscalar and vector charmonium states.
However, we cannot simulate directly at the physical $B$ and $B_c$ masses. In particular, we are limited to heavy-quark masses $am_h < 0.7$ because the residual mass becomes relevant around this region, breaking chiral symmetry explicitly \cite{Cho:2015ffa,Boyle:2018knm}; instead, we study the behavior of the matrix elements of interest for six different values of $am_h$ given in table 2 of ref.~\cite{PhysRevD.109.074503}. This will allow us to extrapolate to the physical heavy-meson masses at a later stage.

For the three-point functions $C_3^{\mu\nu}$, we work in the center-of-mass frame $\pmb{p}=\pmb{0}$ and set the photon momentum with a combination of Fourier modes and twist angles,
$\pmb{k} = 2\pi / L (\pmb{n}-\pmb{\theta_t})$. We keep $\abs{\pmb{n}} \leq 3$ where $\pmb{n}\in\mathds{Z}^3$, and we either employ $\pmb{\theta_t}=\pmb{0}$ or $\pmb{\theta_t}=(0,0,\pm\theta)$ with $\theta=0.2762$. As seen in \cref{fig:modes}, Fourier modes by themselves give us the form factors at several values of $x_\gamma$, but we miss data points for $x_\gamma \sim 0.1-0.2$. This is due to the meson mass and the choice of reference frame. We add one twist angle to remedy this gap in our dataset and increase the density of points at larger values of $x_\gamma$.

We use two types of sources, stochastic $Z_2 \otimes Z_2$ time-wall for the meson interpolator and point sources for the weak current. The meson interpolator is Jacobi smeared,
\begin{equation}
    q(x) \rightarrow \left(1 + \frac{w^2}{4N} \Delta\right)^N q(x)
\end{equation}
with parameters $w=9$ and $N=200$. We form the diagram in \cref{fig:diagram-lattice} in three stages,
\begin{enumerate}
    \item Compute separately the point-to-all quark propagators for every flavour ($ud$, $s$, $h_0$, \dots, $h_5$) and store them on disk. The point-source correlators can have a zero or non-zero twist angle. We locate the meson interpolator on every time-slice and the point sink on every 4th time-slice.
    \item Contract the propagators for every source-sink separation $t_W$, non-zero $(\mu,\nu)$ combination, and value of the twist $\pmb{\theta_t}$.
    Project to all Fourier modes within a sphere of radius $\abs{\pmb{n}} \leq 3$, providing a total of 369 different photon momenta when taking the various twists into account.
\end{enumerate}
From the Lorentz decomposition, it can be seen that there are ten $(\mu,\nu)$ combinations of the vector tensor $H^{\mu\nu}_V$ that vanish and therefore we do not compute: the diagonal $H_V^{\mu\mu}$ is always zero by antisymmetry, and $H_V^{\mu0}$, $H_V^{0\nu}$ vanish in the centre-of-mass frame for a similar reason.

\section{Initial dataset}   \label{sec:dataset}

\begin{figure}
    \centering
    \begin{subfigure}[t]{0.45\textwidth}
        \centering
        \includegraphics[scale=1.3]{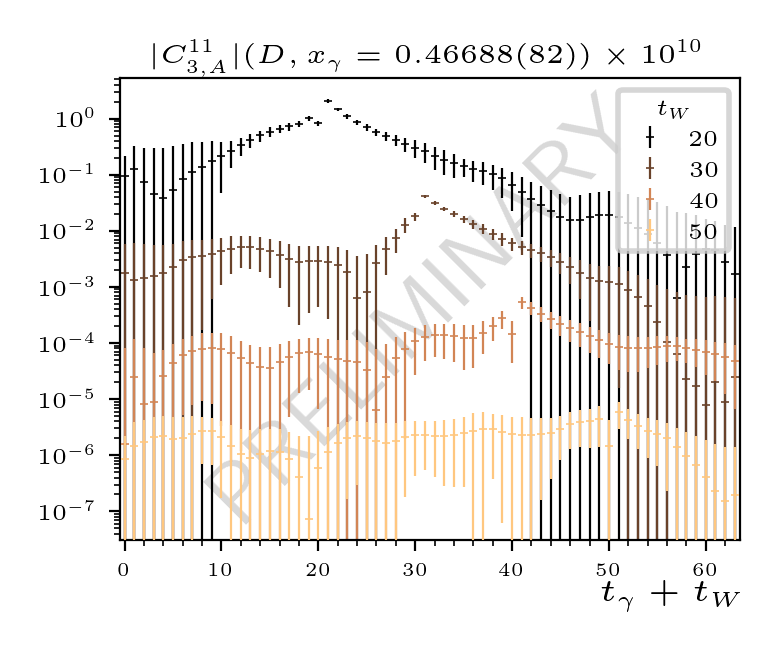}
    \end{subfigure}
    \begin{subfigure}[t]{0.45\textwidth}
        \centering
        \includegraphics[scale=1.3]{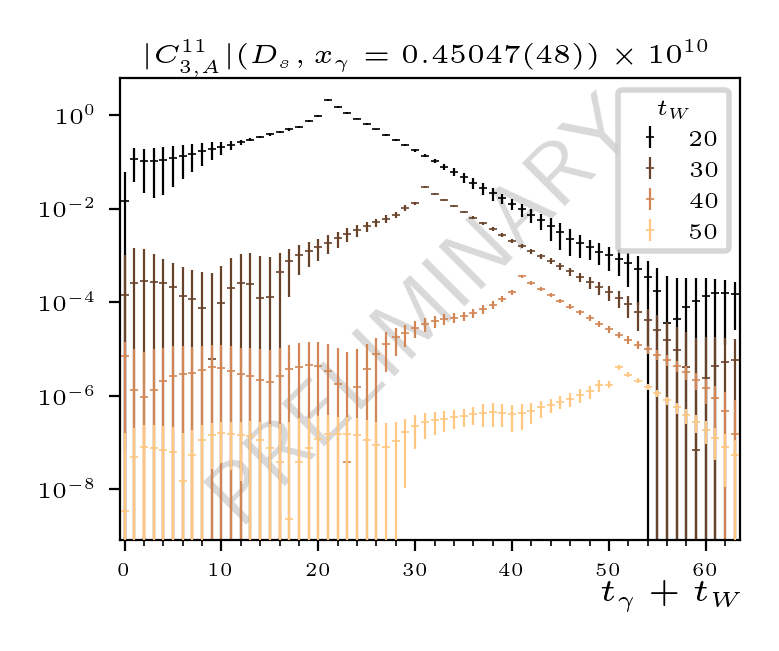}
    \end{subfigure}
    \begin{subfigure}[t]{0.45\textwidth}
        \centering
        \includegraphics[scale=1.3]{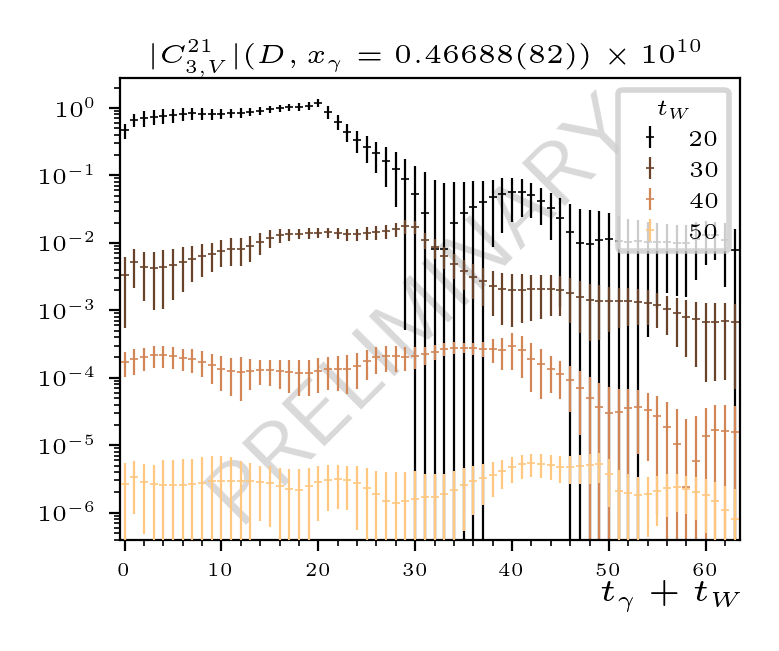}
    \end{subfigure}
    \begin{subfigure}[t]{0.45\textwidth}
        \centering
        \includegraphics[scale=1.3]{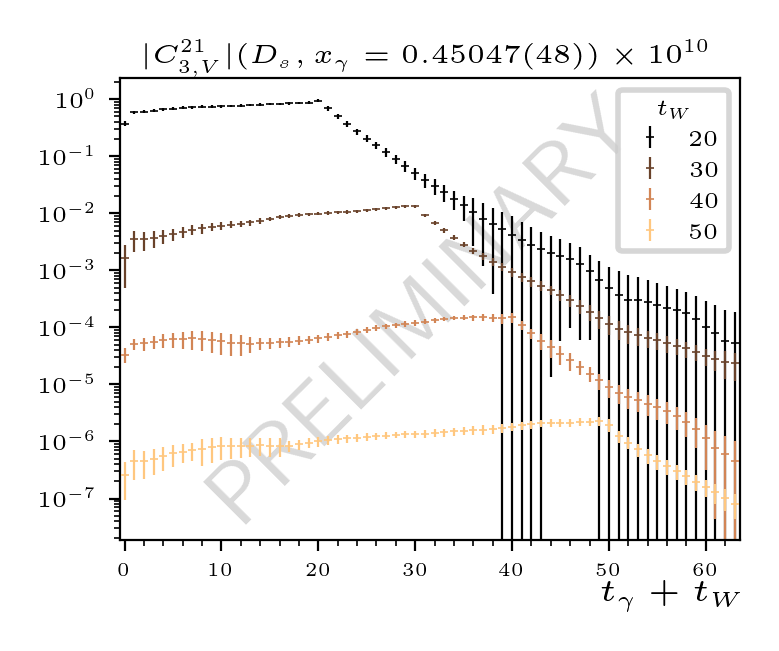}
    \end{subfigure}
    \caption{Axial (top panel) and vector (bottom panel) components of the three-point correlator \cref{eq:3pt} for the Fourier mode $\pmb{n}=(0,0,1)$ and the states $D$ and $D_s$. The top left plot shows the axial component $C_{3,A}^{11}$ for the $D$ meson as a function of the electromagnetic-current time insertion $t_\gamma + t_W$. The four different colors refer to different source-sink separations $t_W$. The other plots correspond to the vector three-point function $C_{3,V}^{21}$ of the $D$ meson (bottom left), the axial $C_{3,A}^{11}$ of the $D_s$ meson (top right), and the vector $C_{3,V}^{21}$ of the $D_s$ meson (bottom right).}
    \label{fig:3pt}
\end{figure}

We have computed the three-point function $C_{3,V-A}^{\mu\nu}$ on 18 gauge configurations of the ensemble shown in \cref{tab:ensembles}, and we measure the correlator at 32 time translations per configuration.
\Cref{fig:3pt} shows the signal for the current statistics in some particular cases. The figure gathers the vector and axial components of the $D$ and $D_s$ three-point correlators with the photon in the $z$ direction and $\pmb{n}=(0,0,1)$. This is indicated by $x_\gamma=2E_\gamma/m_P$ where $E_\gamma$ is the photon energy in the meson rest frame. We plot the absolute value of the correlator as a function of the electromagnetic current time insertion, with the meson interpolator located at $t_\gamma+t_W=0$. The different colors indicate the position of the weak current.

\section{Outlook}   \label{sec:outlook}

With an initial dataset at our disposal, we are ready to study the hadronic tensor $H^{\mu\nu}$. Our method to compute the correlator enables us to study all source-sink separations and non-zero $(\mu,\nu)$ components. To treat excited state contamination, we will test the new Laplace filter technique defined in ref.~\cite{Portelli:2025lop}. In the future, we plan to add data for the quark-disconnected contribution in \cref{fig:quark-disconnected} on the same ensemble. Furthermore, we plan to carry out the calculation in several other lattices at a variety of spacings and quark masses to reach the physical point.

\section{Acknowledgements}
AP would like to warmly thank the RIKEN Centre for Computational Science (R-CCS), where he was hosted for seven months during the initial phase of this project, supported by a Long-Term Invitational Fellowship from the Japanese Society for the Promotion of Science (JSPS).
TSJ, VG, MTH, and AP are supported in part by UK STFC grant ST/X000494/1.
VG, MTH, and AP are additionally supported in part by UK STFC grant ST/T000600/1.
MTH is further supported by UKRI Future Leaders Fellowship MR/T019956/1.
RH is supported in part by the UK STFC grant ST/X000079/1.
NHT is supported by the UK Research and Innovation, Science and Technology Facilities Council, grant number UKRI2426.
MDC has received funding from the European Union’s Horizon Europe research and innovation programme under the Marie Sk\l{}odowska-Curie grant agreement No.\ 101108006.
FE has received funding from the European Union's Horizon Europe research and innovation programme under the Marie Sk\l{}odowska-Curie grant agreement No.\ 101106913.
The work of SH is supported in part by JSPS KAKENHI Grant Number 22H00138, and by the Post-K and Fugaku supercomputer project through the Joint Institute for Computational Fundamental Science (JICFuS).
The work of TK is supported in part by JSPS KAKENHI grant numbers 22K21347, 23K20846 and 25K01007.
The work of YA is supported in part by the project utilizing the Supercomputer Fugaku through the Joint Institute for Computational Fundamental Science (JICFuS) and through System Enhancement and Exploration Category/Usability Research of Fugaku. 
This work used the DiRAC Extreme Scaling service Tursa at the University of Edinburgh, managed by the EPCC on behalf of the STFC DiRAC HPC Facility (www.dirac.ac.uk). The DiRAC service at Edinburgh was funded by BEIS, UKRI and STFC capital funding and STFC operations grants. DiRAC is part of the UKRI Digital Research Infrastructure.

\bibliographystyle{JHEP}
\bibliography{refs}

\end{document}